\documentclass[aps,prl,twocolumn,showpacs,byrevtex]{revtex4}
\let\oldhat\hat
\renewcommand{\hat}[1]{\oldhat{\mathbf{#1}}}

\usepackage{times,mathptm}% for making `pdf' file
\usepackage[dvips]{graphicx,color}% for inclusion of graphics file(s)
\usepackage{titlesec}
\usepackage{array}

\begin{document}
\title{Ferromagnetic Weyl Fermions in Two-Dimensional Layered Electride Gd$_2$C}
\author{Shuyuan Liu$^{1,2}$, Chongze Wang$^{1}$, Liangliang Liu$^3$, Jin-Ho Choi$^2$, Hyun-Jung Kim$^4$, Yu Jia$^3$, Chul Hong Park$^5$, and Jun-Hyung Cho$^{1*}$}
\affiliation{$^1$ Department of Physics, Research Institute for Natural Science, and HYU-HPSTAR-CIS High Pressure Research Center, Hanyang
University, 222 Wangsimni-ro, Seongdong-Ku, Seoul 04763, Republic of Korea \\
$^2$ College of Energy, Soochow Institute for Energy and Materials Innovations and Key Laboratory of Advanced Carbon Materials
and Wearable Energy Technologies of Jiangsu Province, Soochow University, Suzhou 215006, China \\
$^3$ Key Laboratory for Special Functional Materials of the Ministry of Education, Henan University, Kaifeng 475004, People's Republic of China \\
$^4$ Peter Grunberg Institut and Institute for Advanced Simulation, Forschungszentrum J{\"u}lich and JARA, 52425 J{\"u}lich, Germany \\
$^5$ Department of Physics Education, Pusan National University, Pusan 609-735, Republic of Korea}
\date{\today}

\begin{abstract}
Recently, two-dimensional layered electrides have emerged as a new class of materials which possess anionic electron layers in the interstitial spaces between cationic layers. Here, based on first-principles calculations, we discover a time-reversal-symmetry-breaking Weyl semimetal phase in a unique two-dimensional layered ferromagnetic (FM) electride Gd$_2$C. It is revealed that the crystal field mixes the interstitial electron states and Gd 5$d$ orbitals near the Fermi energy to form band inversions. Meanwhile, the FM order induces two spinful Weyl nodal lines (WNLs), which are converted into multiple pairs of Weyl nodes through spin-orbit coupling. Further, we not only identify Fermi-arc surface states connecting the Weyl nodes but also predict a large intrinsic anomalous Hall conductivity due to the Berry curvature produced by the gapped WNLs. Our findings demonstrate the existence of Weyl fermions in the room-temperature FM electride Gd$_2$C, therefore offering a new platform to investigate the intriguing interplay between electride materials and magnetic Weyl physics.
\end{abstract}
\maketitle
%\begin{multicols}{2}

%\vspace{0.4cm}
%\section{I. INTRODUCTION}
%\vspace{0.4cm}

In the past decade, two-dimensional (2D) layered materials have attracted tremendous attention because of their promising opportunities in both fundamental research and technological applications~\cite{Graphene-Science2004,Graphene-Rev2009,TMDCs-Rev2017,2D-herer.-Science2016}. Among them, 2D layered electrides~\cite{Electride-PRX2014,Electride-Y2C2014,Electride-JACS2016,Seho} are highly attractive as a novel class of unconventional ionic compounds, where some excess electrons are confined in the interstitial spaces between positively charged cationic layers. These interstitial electrons behaving as interstitial quasiatoms~\cite{Electride-Hoffman-JACS2014} or anions form a 2D electron gas with exciting electronic properties, such as high electrical conductivities, low work functions, highly anisotropic optical response, and rich surface-related chemical reactions~\cite{Ga2N-Nature2013,Dye-Science2013,catalysis-Nat.Chem2012}. Interestingly, depending on the cationic constituent atoms, interstitial electrons in 2D layered electrides have different degrees of localization to form various electronic phases~\cite{shuyuan-jpcc}. For example, the 2D layered electrides of Ca$_2$N~\cite{Ga2N-Nature2013} and Gd$_2$C~\cite{Gd2C-Nat.Commun.2020} which show the features of delocalized and localized interstitial electrons [see Fig. 1(a)] have been experimentally and theoretically demonstrated to exhibit the nonmagnetic (NM) and ferromagnetic (FM) phases, respectively.

\begin{figure}[h!t]
\includegraphics[width=8.5cm]{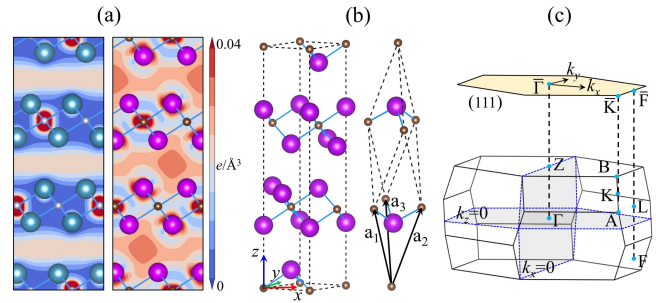}
\caption{(a) Calculated partial charge densities of Ca$_2$N (left) and Gd$_2$C (right), (b) optimized structure of Gd$_2$C within the trigonal conventional cell (left) and rhombohedral primitive cell (right), and (c) Brillouin zone of the rhombohedral primitive cell. In (a), each charge density integrated between $E_{\rm F}-$1.0 eV and $E_{\rm F}$ is plotted on the (110) plane of the conventional unit cell. The large and small spheres represent Ca (Gd) cations and N (C) anions, respectively. In (c), the $k_z$ = 0 and $k_x$ = 0 planes are drawn, and the (111) surface Brillouin zone is also drawn.}
\label{figure:1}
\end{figure}

Because of the loosely bound nature of interstitial electrons away from the cationic layers, 2D layered electrides tend to have the corresponding bands near the Fermi energy $E_{\rm F}$~\cite{Ga2N-Nature2013}. These electronic properties of 2D layered electrides could be favorable for achieving topological phases through band inversions~\cite{Bradlyn2017,Po2017}, which are characterized by the nontrivial topology of bulk bands and its associated robust surface states. The resulting combination between electride property and topological electronic characteristics provides versatile platforms to design electrides with new functionalities that offer great promises in future electronic device applications. Recently, it has been suggested that some 2D layered electrides host nontrivial topological phases such as topological insulators Sc$_2$C and HfBr as well as nodal-line semimetals Y$_2$C, Sr$_2$Bi, and Rb$_3$O~\cite{FengLiu,LiangLiu,Topological-Electride-PRX2018,Topological-Electride-PRM2019,Topological-Electride-JPCC2019}. All of these topological phases arising from the NM or paramagnetic electrides limit their tunability and functionality with magnetization. On the other hand, the 2D layered FM electride Gd$_2$C with a high Curie temperature of ${\sim}$350 K~\cite{Gd2C-Nat.Commun.2020} is here demonstrated to have a novel quantum state of magnetic Weyl semimetal (WSM), which hosts Weyl fermions in the bulk and unique Fermi-arc surface states connecting pairs of Weyl nodes. Indeed, magnetic WSMs have a variety of novel magnetic responses such as chiral magnetic effects, large anomalous Hall conductivity, and quantum anomalous Hall effect~\cite{Spintronic-PRAppl2016,Spintronic-Nat.Phys.2017}. In contrast to many inversion symmetry-broken WSMs~\cite{TaAs-PRX2015,TaAs-Nat.Phys.2015,WTe2-Nature2015,MoTe2-PRX2016}, magnetic WSMs with breaking time-reversal symmetry (TRS) have been experimentally realized in only a few ternary nonelectride compounds such as GdPtBi~\cite{GdPtBi}, Y$_2$Ir$_2$O$_7$~\cite{Y2Ir2O7}, HgCr$_2$Se$_4$~\cite{HgCr2Se2,HgCr2Se2-Expt}, and Co$_3$Sn$_2$S$_2$~\cite{Co3Sn2S2-Nat.Phy.2018,Co3Sn2S2-Science2019-1,Co3Sn2S2-Science2019-2}. Therefore, as the first FM WSM electride, Gd$_2$C is the unique material involving the interplay of electride nature, ferromagnetism, and nontrivial band topology.

In this Letter, using first-principles calculations, we discover a WSM phase in the FM electride Gd$_2$C that was recently synthesized ~\cite{Gd2C-Nat.Commun.2020}. This TRS-breaking WSM phase is driven by the combined effects of crystal field, ferromagnetism, and spin-orbital coupling (SOC). We find that the crystal field inverts the hybridized bands of the interstitial electron states and Gd 5$d$ orbitals near $E_{\rm F}$, which lead to the formation of band crossings. Meanwhile, the FM order induces spinful Weyl nodal lines (WNLs) having twofold degenerate spin-up or spin-down band crossings along one-dimensional lines in momentum space. Finally, the inclusion of SOC transforms the WNLs into multiple pairs of Weyl nodes. We further demonstrate the existence of topological Fermi-arc states on the (111) surface as well as strong intrinsic anomalous Hall effects due to the large Berry curvature created by the SOC-driven gap openings of WNLs. The present study has important implications for understanding the FM Weyl physics involved in 2D layered electrides such as Gd$_2$C as well as other isostructural lanthanide carbides Tb$_2$C and Dy$_2$C, which will stimulate experimental studies in the future.

\begin{figure}[h!t]
\includegraphics[width=8.5cm]{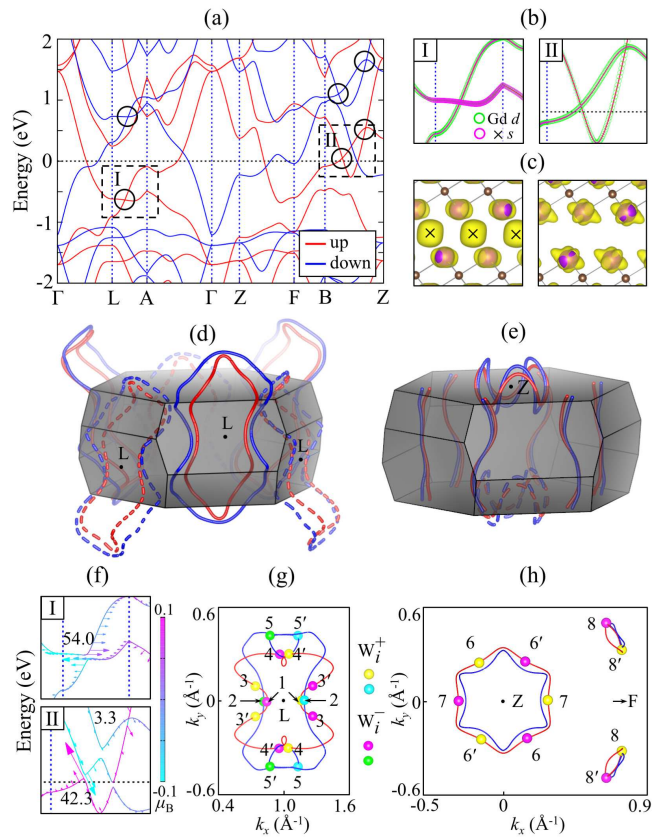}
\caption{(a) Calculated band structure of Gd$_2$C using the DFT + U calculation in the absence of SOC, (b) projected bands onto the Gd-$d$ and interstitial-$s$-like orbitals in the regions I and II, (c) charge character of the Weyl node in the region I with an isosurface of 0.03 $e$/{\AA}$^3$, and (d),(e) spinful WNLs. The energy zero in (a) represents $E_{\rm F}$. The radii of circles in (b) are proportional to the weights of the corresponding orbitals. The dashed lines in (d) and (e) represent the identical WNLs with modulo the corresponding reciprocal vectors. The closeup band structures in the regions I and II, computed with including SOC, are displayed in (f), where the arrows represent the $m_x$ (horizontal direction) and $m_y$ (vertical direction) components and the arrow colors indicate $m_z$. The values $m_x$, $m_y$, and $m_z$ for the longest arrow in I (II) are $-$0.252, 0, and $-$0.123 ($-$0.187, 0.324, and 2.021) ${\mu}_{\rm B}$, respectively. The numbers in (f) represent the SOC-induced gap size in meV. The $k_x$ and $k_y$ positions of the Weyl nodes existing in the energy range between $E_{\rm F}-$1.0 and $E_{\rm F}+$1.0 eV are drawn in (g) and (h).
}
\label{figure:2}
\end{figure}

We first optimize the structure of Gd$_2$C using the DFT + U calculation without including SOC~\cite{method}. Figures 1(b) and 1(c) show the optimized structure with the lattice parameters $a_1$ = $a_2$ = $a_3$ = 6.516 {\AA} and its Brillouin zone, respectively. The calculated spin-up and spin-down bands are displayed in Fig. 2(a). We find that there are two (four) spinful linear band crossings along the $\overline{\rm LA}$ ($\overline{\rm BZ}$) line around $E_{\rm F}$, indicated by circles in Fig. 2(a). The linear crossings along the $\overline{\rm LA}$ line continuously evolve to form three-dimensional (3D) closed-loop shapes of WNLs around the $L$ points [see Fig. 2(d)]. Meanwhile, those along the $\overline{\rm BZ}$ line form two species of WNLs: one is the closed loops around the $Z$ point and the other is the open lines along the $k_z$ direction [see Fig. 2(e)]. It is noted that the spin-up and spin-down bands are decoupled with each other in the absence of SOC. Since the crystalline symmetries of Gd$_2$C belong to the space group $R\overline{3}m$ (No. 225) with the point group $D_{3d}$ containing space-inversion symmetry $P$, threefold rotational symmetry $C_{3z}$ about the $z$ axis, and twofold rotation symmetry $C_{2}$ about the $y$ axis and two $y'$ axes (obtained by the ${\pm}$120$^{\circ}$ rotation of the $y$ axis about the $z$ axis), we find a total of twenty WNLs in the whole Brillouin zone, as shown in Fig. 2(d) and 2(e). To identify the electronic origin of WNLs, we project the crossing bands onto the Gd-5$d$ and interstitial-$s$-like orbitals [see Fig. 2(b)], which are more dominant components compared to other orbitals (see Fig. S1 in the Supplemental Material~\cite{SM}). It is thus likely that the band inversions along the $\overline{\rm LA}$ and $\overline{\rm BZ}$ lines are derived from the Gd-5$d$ and interstitial-$s$-like orbitals. As shown in Fig. 2(c), the charge character of the doubly degenerate Weyl node in the region I represents major distributions at Gd atoms and the interstitial regions ${\times}$ between interlayers. It was reported~\cite{FengLiu} that the NM phase of Y$_2$C has a closed-loop shape of Dirac nodal lines (DNLs) around the $L$ points with a hybridization of the Y-4$d$ and interstitial-$s$-like orbitals. Therefore, we can say that the crystal fields of isostructural Y$_2$C and Gd$_2$C with the same crystalline symmetries similarly induce band inversions/crossings near $E_{\rm F}$, leading to their DNLs and WNLs, respectively.

In contrast to usual nodal lines residing on 2D mirror planes of the Brillouin zone~\cite{2Dnodal-line}, the present WNLs exhibit the 3D snakelike wrapped loops or lines [see Figs. 2(d) and 2(e)]. We note that the DNLs in Y$_2$C are protected by $P$ symmetry and TRS~\cite{FengLiu}. Meanwhile, the FM order of Gd$_2$C breaking TRS produces two spinful WNLs, which are protected by $P$ symmetry. To confirm the symmetry protection of the WNLs, we introduce various displacements of Gd atoms which break or preserve $P$ symmetry. Obviously, our calculated DFT band show the gap openings of WNLs for the $P$-symmetry-breaking geometries (see Fig. S2 in the Supplemental Material~\cite{SM}). We further identify the nontrivial topological characterization of WNLs by calculating the topological index~\cite{Z2index}, defined as ${\zeta}_1$ = ${\frac{1}{\pi}}$ ${\oint}$$_c$ $dk$${\cdot}$A($k$), along a closed loop encircling any of the present WNLs. Here, A(k) = $-i$$<$$u_k$$\mid$$\partial$$_k$$\mid$$u_k$$>$ is the Berry connection of the related Bloch bands. We obtain ${\zeta}_1$ = ${\pm}$1 for the WNLs, indicating that they are stable against the lattice deformations conserving $P$ symmetry.

Next, we calculate the band structure of Gd$_2$C using the DFT + U calculation with including SOC. We find that SOC lifts the degeneracy at the band-crossing points along the $\overline{\rm LA}$ and $\overline{\rm BZ}$ lines [see Fig. 2(f)]. Each spin-up (spin-down) WNL around the $L$ points becomes gapped with the exception of five (three) pairs of Weyl nodes, while the two species of spin-up (spin-down) WNLs in Fig. 2(e) are converted into three and two (six and two) pairs of Weyl nodes. Figures 2(g) and 2(h) display the distributions of Weyl nodes within the energy range between $E_{\rm F}-$1.0 and $E_{\rm F}$+1.0 eV. Since spin is no longer a good quantum number in the presence of SOC, three additional pairs of Weyl nodes between the spin-up and spin-down bands appear around the $Z$ point (see Fig. S3 in the Supplemental Material~\cite{SM}). We note that the magnitudes of the SOC-driven gaps of WNLs are determined by orbital magnetic moments [see Fig. 2(f)], which are generated by the hybridization of the Gd-5$d$ and interstitial-$s$-like orbitals [see Fig. 2(b)]. As a consequence of SOC, the spontaneous magnetization points along the $z$ axis, which is more favorable than the magnetization direction along the $x$ axis by 0.12 meV per primitive unit cell (see Table SI in the Supplemental Material~\cite{SM}). Therefore, the magnetic point group of the system changes into C$_{3i}$ containing $P$, $C_{3z}$, and the product $C_{2}T$ of rotation $C_{2}$ and TRS $T$. As shown in Figs. 2(f) and 2(g), each pair of Weyl nodes associated with inversion $P$ has its counterpart through $C_{2y}T$ symmetry, e.g., (W$_{3}^{+}$, W$_{3}^{-}$) and (W$_{3'}^{+}$, W$_{3'}^{-}$), while the counterparts of (W$_{1}^{+}$, W$_{1}^{-}$), (W$_{2}^{+}$, W$_{2}^{-}$), and (W$_{7}^{+}$, W$_{7}^{-}$) are themselves. Here, the chiral charge of each Weyl node is determined by calculating the Berry flux through a 2D closed surface enclosing the node. The computed chiral charges (i.e., Chern numbers) have the values of +1 and $-$1 for W$_{i}^{+}$ and W$_{i}^{-}$, respectively. The positions of all twenty-four pairs of Weyl nodes in the Brillouin zone are given in Table SII of the Supplemental Material~\cite{SM}, together with their energies.

\begin{figure}[h!t]
\includegraphics[width=8.5cm]{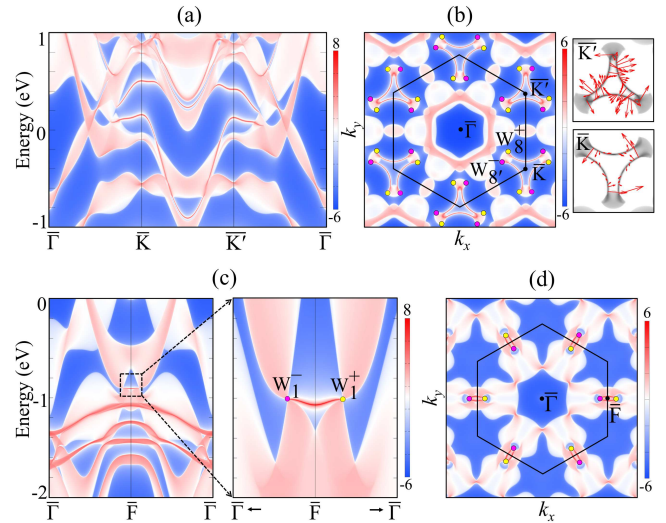}
\caption{(a) Projected surface spectrum for the (111) surface of Gd$_2$C with (b) the isoenergy surface at 0.06 eV above $E_{\rm F}$. The
in-plane spin textures of the surface states around the $\overline{\rm K}$ and $\overline{\rm K'}$ points are also drawn in (b). Another projected surface spectrum along the $\overline{\Gamma}-\overline{\rm F}-\overline{\Gamma}$ and the isoenergy surface at $-$0.90 eV below $E_{\rm F}$ are displayed in (c) and (d), respectively. A closeup image around the $\overline{\rm F}$ point is given in (c).}
\label{figure:3}
\end{figure}

To elucidate the existence of topologically protected surface states that represent the essential hallmark of Weyl nodes, we calculate the surface electronic structure of Gd$_2$C using the Green's function method based on the tight-binding Hamiltonian with maximally localized Wannier functions~\cite{wannnier90,wanniertools}. Figure 3(a) shows the projected surface spectrum for the (111) surface of Gd$_2$C. We find topological surface states around the $\overline{\rm K}$ and $\overline{\rm K'}$ points near $E_{\rm F}$. Such nontrivial surface states can be more distinguishable by subtracting the (111) projected bulk states from the surface spectrum (see Fig. S4 in the Supplemental Material~\cite{SM}). In Fig. 3(b), we plot the Fermi surface of the (111) surface at a chemical potential of 0.06 eV above $E_{\rm F}$. It is obviously seen that there are the open Fermi arcs connecting W$_{8}^{+}$(W$_{8'}^{+}$) and W$_{8'}^{-}$ (W$_{8}^{-}$) around the $\overline{\rm K}$ ($\overline{\rm K'}$) point. Furthermore, these topological surface states are spin-polarized with helical spin textures [see the insets in Fig. 3(b)], reflecting a unique spin-momentum locking property in topological surface states~\cite{helicalspin}. Figure 3(c) also shows the topological surface states connecting W$_{1}^{+}$ and W$_{1}^{-}$ around the $\overline{\rm F}$ point. Interestingly, the Fermi surface at a chemical potential of $-$0.90 eV below $E_{\rm F}$ exhibits two Fermi arcs connecting the W$_{1}^{+}$ and W$_{1}^{-}$ nodes, similar to the drumheadlike surface states obtained without including SOC (see Fig. S5 of the Supplemental Material~\cite{SM}). This tiny variation of topological surface states with including SOC is likely caused by small energy gap openings along the WNL connecting W$_{1}^{+}$ and W$_{1}^{-}$.

The search for room-temperature magnetic WSM materials has lately been very demanding of attention because of their novel magnetic responses that may be used for future spintronics technologies~\cite{Co3Sn2S2-Science2019-1,Co3Sn2S2-Science2019-2,Co2MnGa}. In order to estimate the Curie temperature $T_{\rm c}$ of Gd$_2$C, we perform spin-polarized calculations for the FM and three different antiferromagnetic (AFM) configurations (see Fig. S6 in the Supplemental Material~\cite{SM}). The FM configuration is found to be more stable than the lowest-energy AFM configuration by 48.6 meV per Gd atom. Using the mean field approximation~\cite{MFA}, we estimate a $T_{\rm c}$ of ${\sim}$377 K, in good agreement with the experimental value of $T_{\rm c}$ ${\approx}$ 350 K~\cite{Gd2C-Nat.Commun.2020}. The resulting room-temperature FM WSM phase is expected to exhibit the intrinsic anomalous Hall effect originating from Berry curvature. Figure 4(a) shows the energy-dependent anomalous Hall conductivity ${\sigma}_{xy}$ calculated by integrating the Berry curvature along $k_z$ in the Brillouin zone. We find that ${\sigma}_{xy}$ is 399 ${\Omega}^{-1}$cm$^{-1}$ at $E_{\rm F}$, while it shows a large peak of ${\sim}$1167 ${\Omega}^{-1}$cm$^{-1}$ at 0.06 eV above $E_{\rm F}$. It is thus likely that an electron doping of ${\sim}$0.07$e$ per primitive unit cell shifting $E_{\rm F}$ to 0.06 eV (see Fig. S7 in the Supplemental Material~\cite{SM}) could increase ${\sigma}_{xy}$ significantly. Figures 4(b) and 4(c) show the Berry curvature distributions in the $k_z$ = 0 and $k_x$ = 0 planes [see Fig. 1(c)], respectively. We find that there are hot spots of the Berry curvature, originating from the gapped WNLs around W$_{8}$, W$_{13}$, and W$_{14}$ (see Table SII in the Supplemental Material~\cite{SM}). Such intrinsic Berry curvature gives rise to the large peak of ${\sigma}_{xy}$ that is a fascinating electronic transport behavior.

\begin{figure}[h!t]
\includegraphics[width=8.5cm]{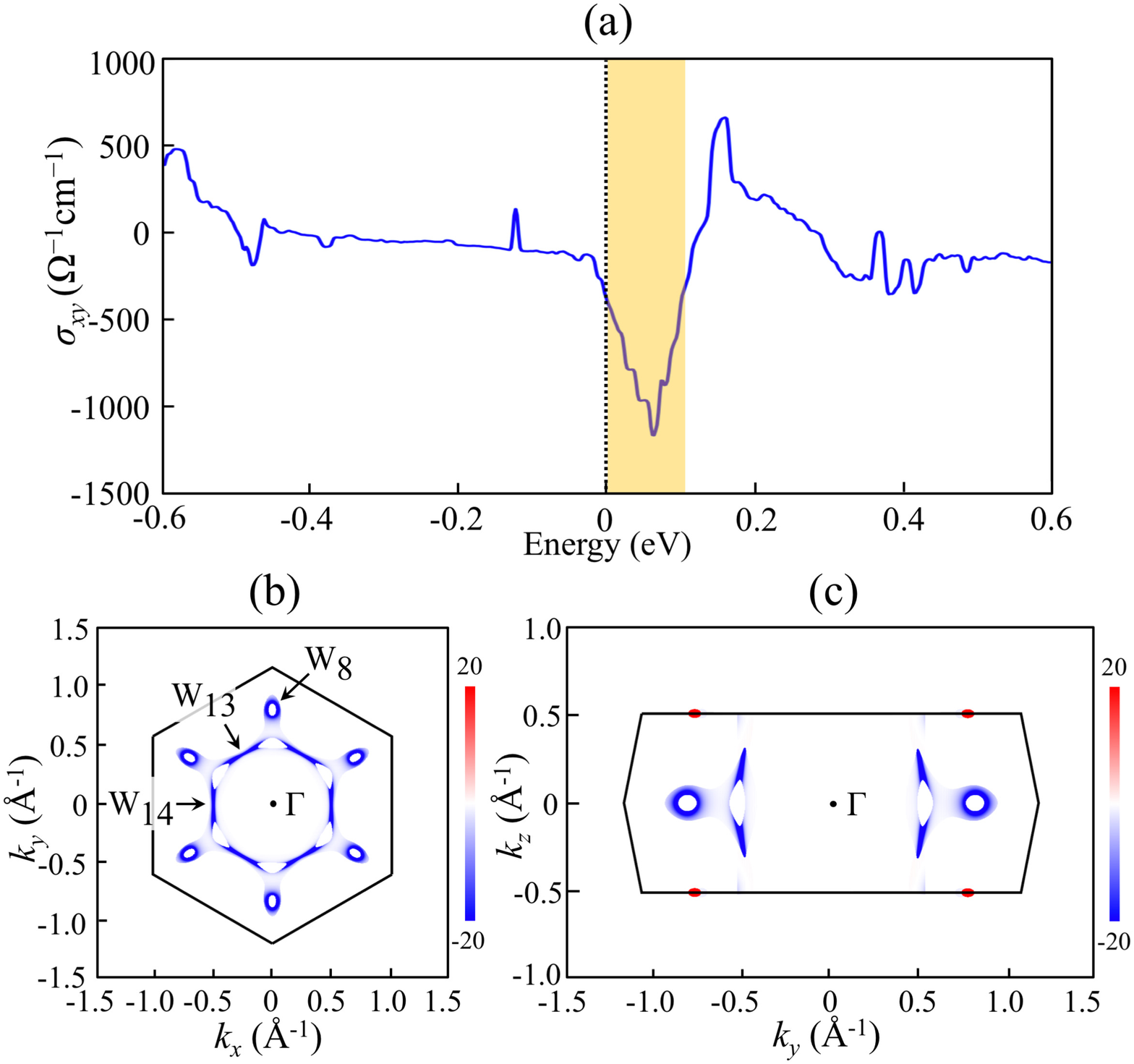}
\caption{(a) Energy dependence of the anomalous Hall conductivity ${\sigma}_{xy}$ and Berry curvature distribution in (b) the $k_z$ = 0 plane and (c) the $k_x$ = 0 plane of the Brillouin zone [see Fig. 1(c)]. In (b) and (c), the $z$ component of Berry curvature is integrated between $E_{\rm F}$ and $E_{\rm F}$+0.1 eV [see the yellow colored region in (a)], and the red and blue regions indicate positive and negative values of Berry curvature, respectively. }
\label{figure:3}
\end{figure}

To conclude, based on first-principles calculations, we have discovered the emergence of WSM phase in a 2D layered FM electride material Gd$_2$C. We revealed that this TRS-breaking WSM phase is driven by the combined effects of crystal field, ferromagnetism, and SOC. The crystal field inverts the hybridized bands of the interstitial electron states and Gd 5$d$ orbitals near $E_{\rm F}$, and the FM order induces two spinful WNLs with the shapes of 3D snakelike wrapped loops and lines. Finally, SOC splits the WNLs into multiple pairs of Weyl nodes with opposite chirality. We further demonstrated not only Fermi-arc surface states connecting the Weyl nodes but also large intrinsic anomalous Hall conductivity originating from the Berry curvature created by the gapped WNLs. It is noteworthy that other isostructural lanthanide carbides such as Tb$_2$C and Dy$_2$C also show similar band dispersions and band crossings near $E_{\rm F}$ (see Fig. S8 in the Supplemental Materia~\cite{SM}), thereby hosting magnetic WSM phases. Therefore, our results not only establish the intriguing interplay of electride properties, ferromagnetism, and nontrivial Weyl band topology in Gd$_2$C, but also open the door for the emergence of topological WSM phase in 2D layered FM lanthanide carbides.

\vspace{0.4cm}

\noindent {\bf Acknowledgements.}
This work was supported by the National Research Foundation of Korea (NRF) grant funded by the Korean Government (Grants No. 2019R1A2C1002975, No. 2016K1A4A3914691, and No. 2015M3D1A1070609). The calculations were performed by the KISTI Supercomputing Center through the Strategic Support Program (Program No. KSC-2019-CRE-0183) for the supercomputing application research.  \\

S. L. and C. W. contributed equally to this work.

%\vspace{0.4cm}
%\noindent {\bf Acknowledgement.}
%\vspace{0.4cm}

                  %%%%%  REFERENCES  %%%%%
\noindent $^{*}$ Corresponding author: chojh@hanyang.ac.kr

\newpage
\onecolumngrid
\newpage
\titleformat*{\section}{\LARGE\bfseries}

\renewcommand{\thefigure}{S\arabic{figure}}
\setcounter{figure}{0}

\vspace{1.2cm}

\section{Supplemental Material for “Ferromagnetic Weyl Fermions in Two-Dimensional Layered Electride Gd$_2$C”}
\vspace{1cm}
\begin{flushleft}

{\bf 1. Band projections onto the Gd, C, and X orbitals}
\begin{figure}[ht]
\includegraphics[width=15cm]{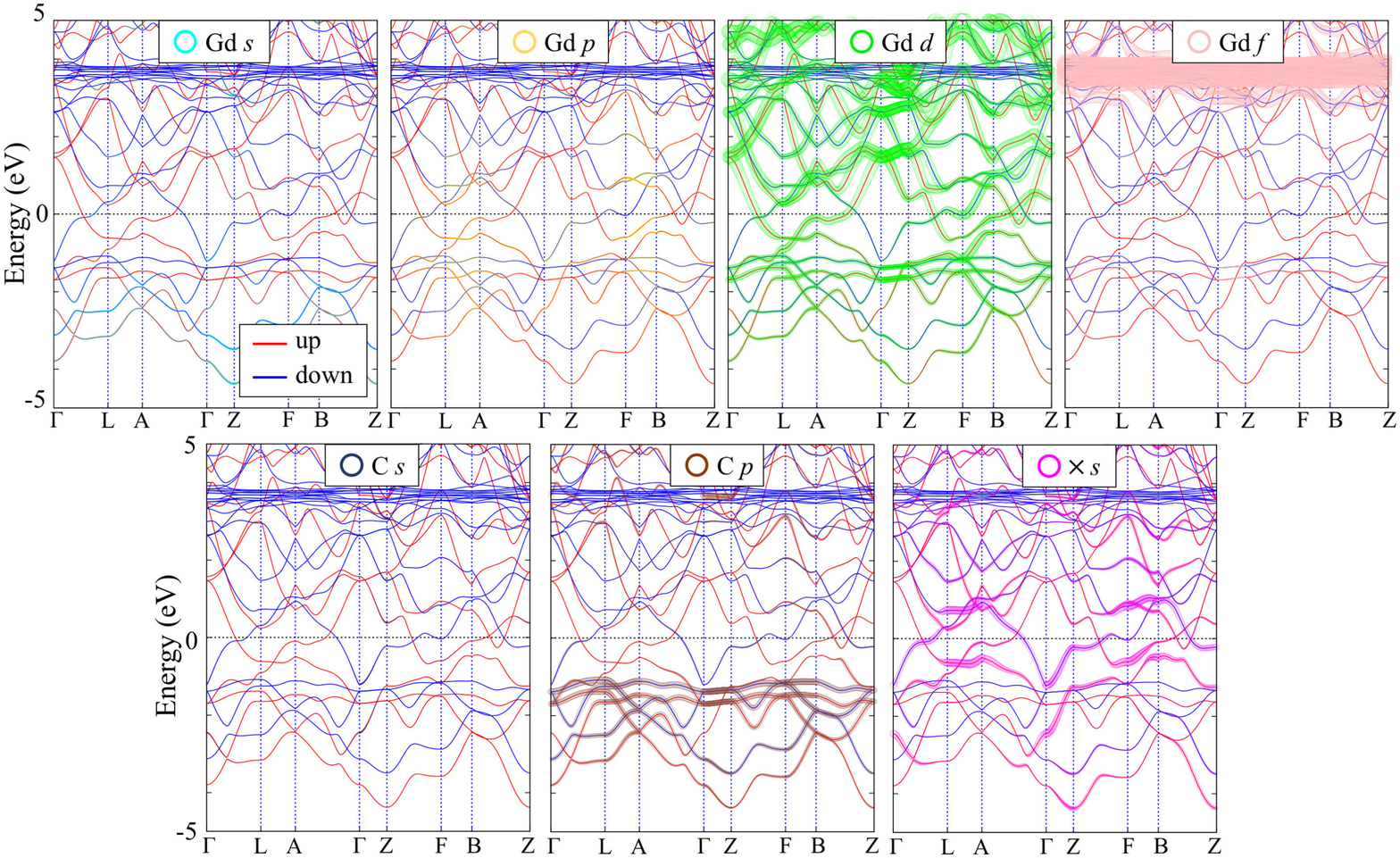}
\caption{Calculated bulk Gd$_2$C bands projected onto the Gd $s$, Gd $p$, Gd $d$, Gd $f$, C $s$, C $p$, and X $s$ orbitals. Here the radii of circles are proportional to the weights of the corresponding orbitals. The dominant components of the WNLs near the Fermi energy are the Gd $d$ and X $s$ orbitals, compared to other orbitals.}
\end{figure}

\vspace{1.2cm}

{\bf 2. Structure of breaking inversion symmetry and its band structure}
\begin{figure}[ht]
\includegraphics[width=11cm]{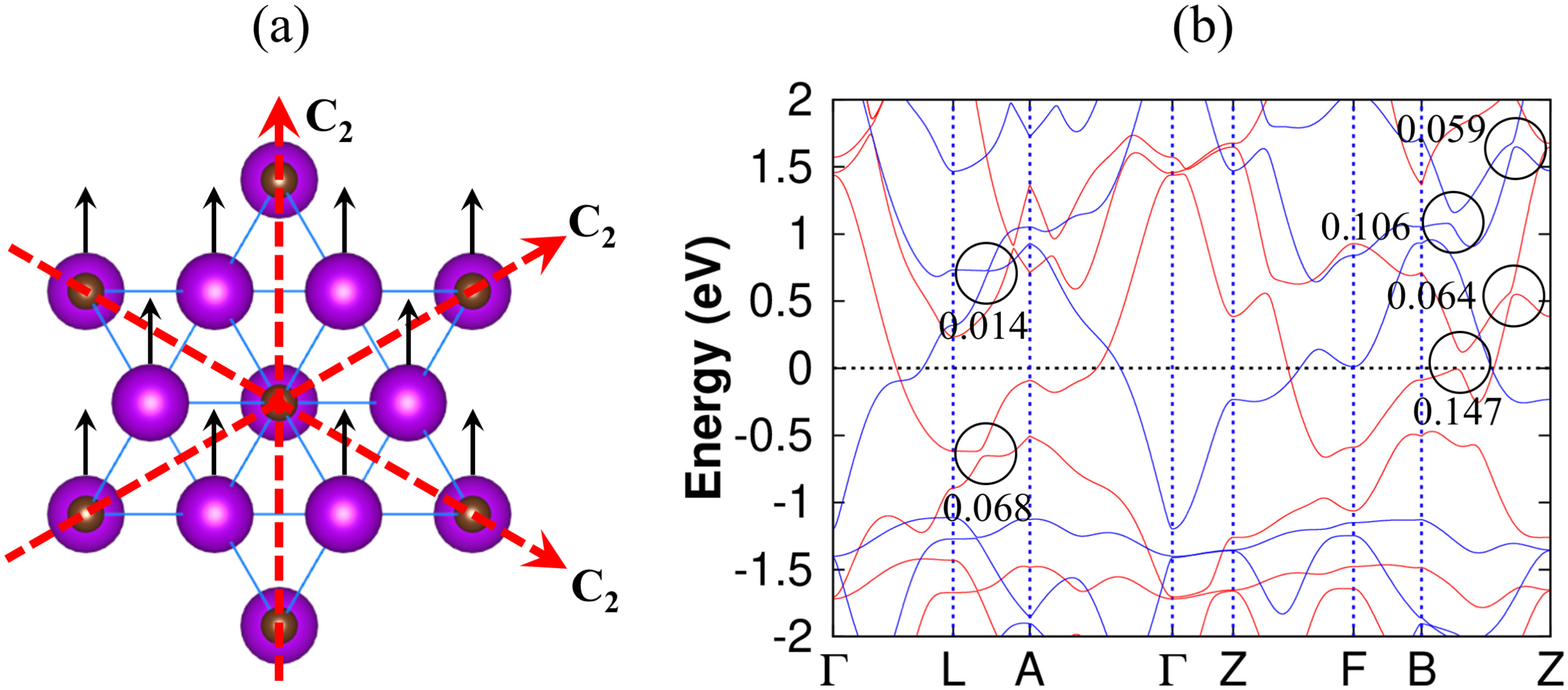}
\caption{(a) Bulk Gd$_2$C structure with broken inversion symmetry and (b) its band structure. Here, the arrows indicate the 0.1 {\AA} shifts of Gd atoms from the optimized structure. The numbers in (b) represent the gap openings (in the unit of eV) at the crossing points.}
\end{figure}

\newpage

{\bf 3. Band structure of Gd$_2$C obtained using the PBE+U calculation with including SOC }
\begin{figure}[ht]
\includegraphics[width=10.9cm]{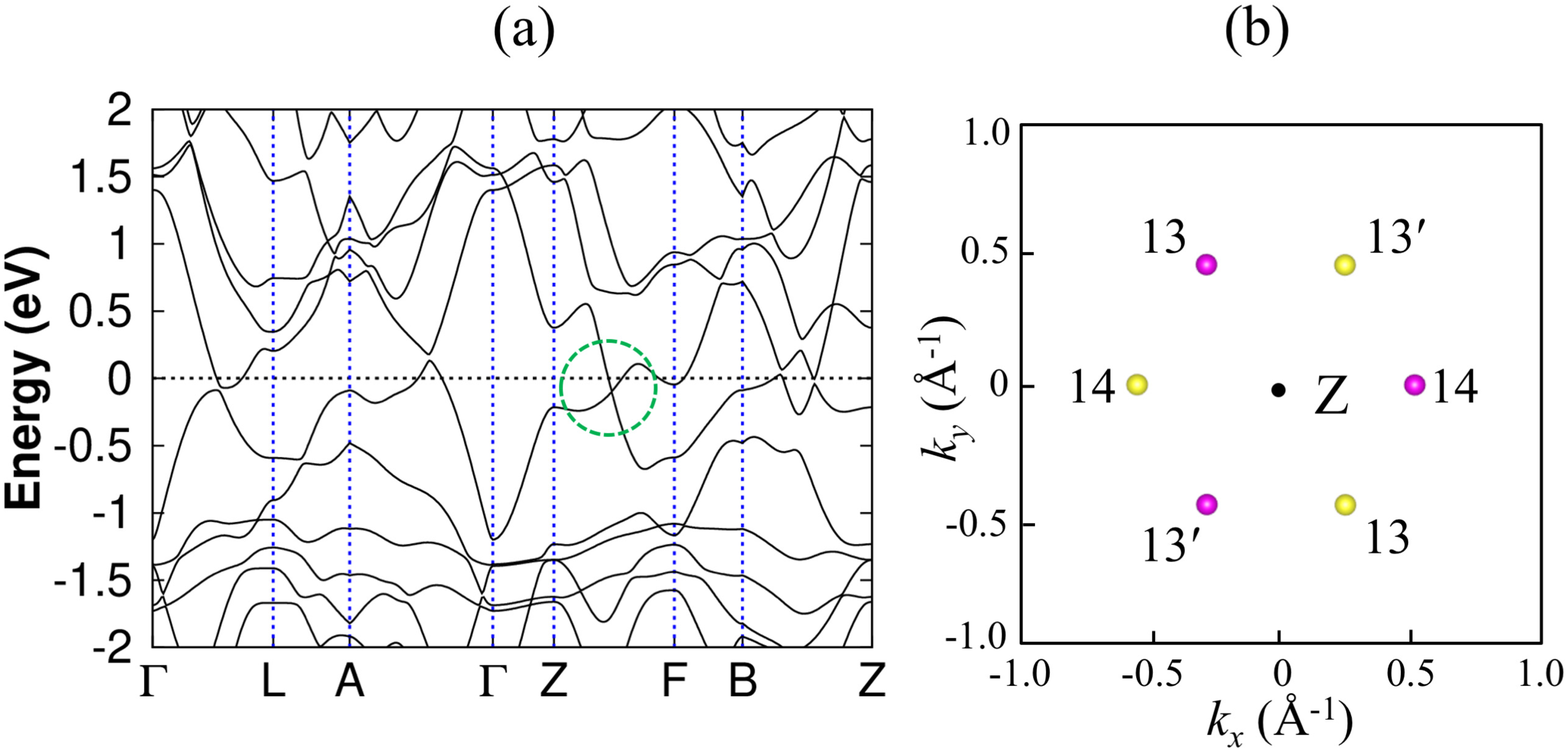}
\caption{(a) Band structure of Gd$_2$C obtained using the PBE+U calculation with including SOC and (b) additional Weyl nodes. In (a), the circle represents the band crossing between the spin-up and spin-down bands along the Z-F line. The $k_x$ and $k_y$ positions of three additional Weyl nodes around the Z point are drawn in (b).}
\end{figure}

\vspace{0.4cm}

{\bf 4. Difference between surface and bulk spectra for the Gd$_2$C(111) surface}
\begin{figure}[ht]
\includegraphics[width=13.5cm]{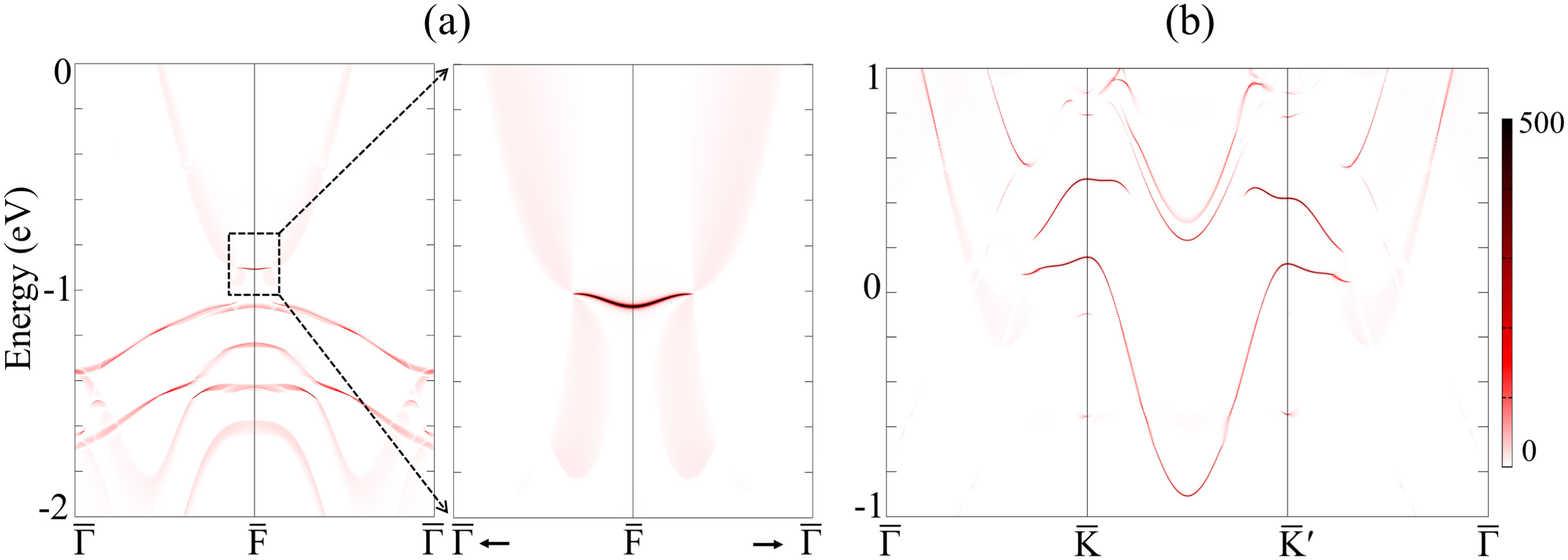}
\caption{Calculated difference between surface and bulk band spectra for the (111) surface of Gd$_2$C along the (a) $\overline{\Gamma}-\overline{\rm F}-\overline{\Gamma}$ and (b) $\overline{\Gamma}-\overline{\rm K}-\overline{\rm K'}-\overline{\Gamma}$ lines. The nontrivial surface states are evidenced by the difference between the surface bands and the (111) projected bulk bands.}
\end{figure}

\vspace{0.4cm}

{\bf 5. Drumheadlike surface state in the absence of SOC}
\begin{figure}[ht]
\includegraphics[width=5.2cm]{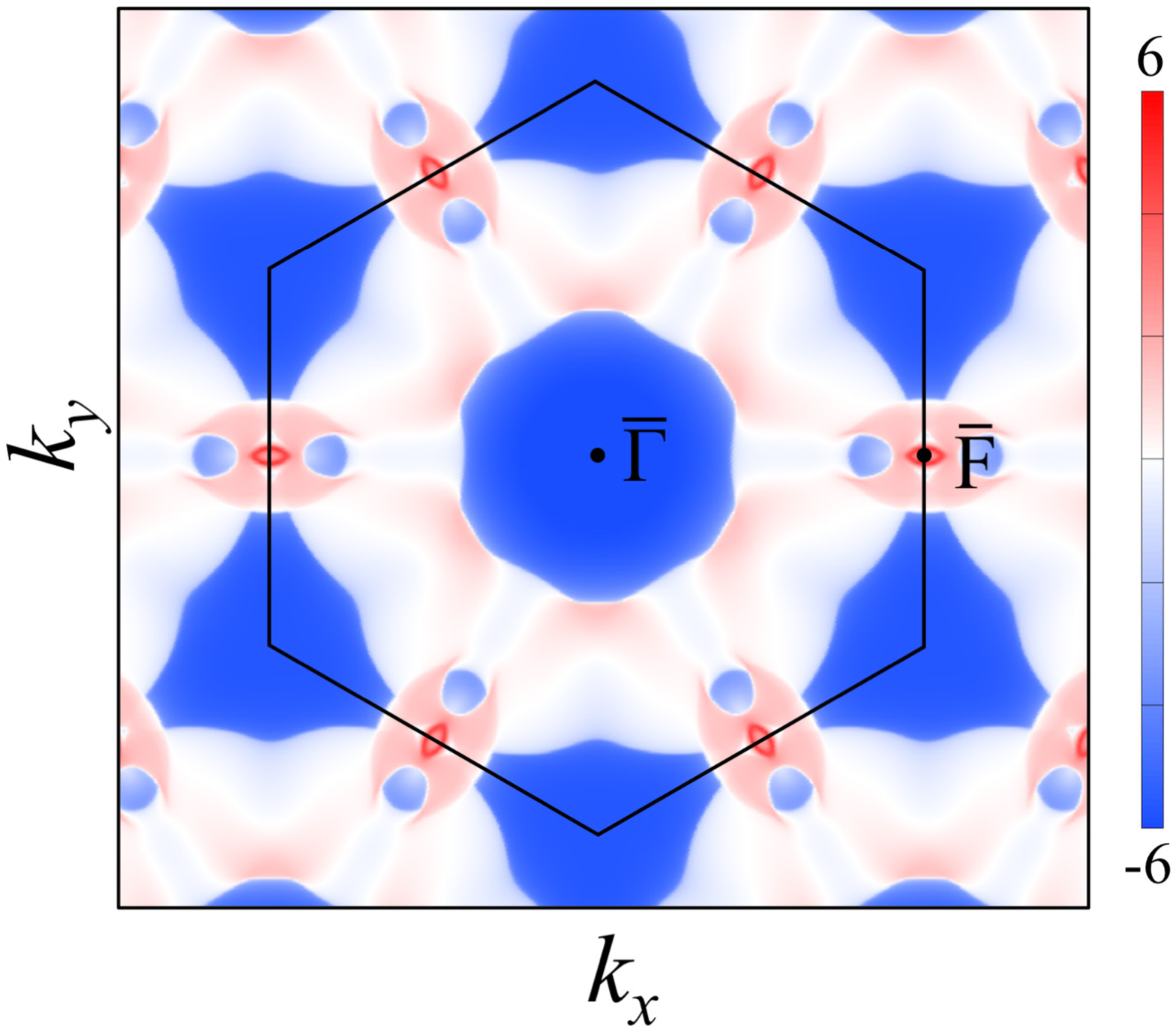}
\caption{Projected Fermi surface of the WNL semimetal state without including SOC, obtained at a chemical potential of $-$0.90 eV below $E_{\rm F}$. This isoenergy surface shows the shape of the drumheadlike surface state.}
\end{figure}

\newpage

{\bf 6. Magnetic structures and Curie temperature of Gd$_2$C }
\begin{figure}[ht]
\includegraphics[width=13cm]{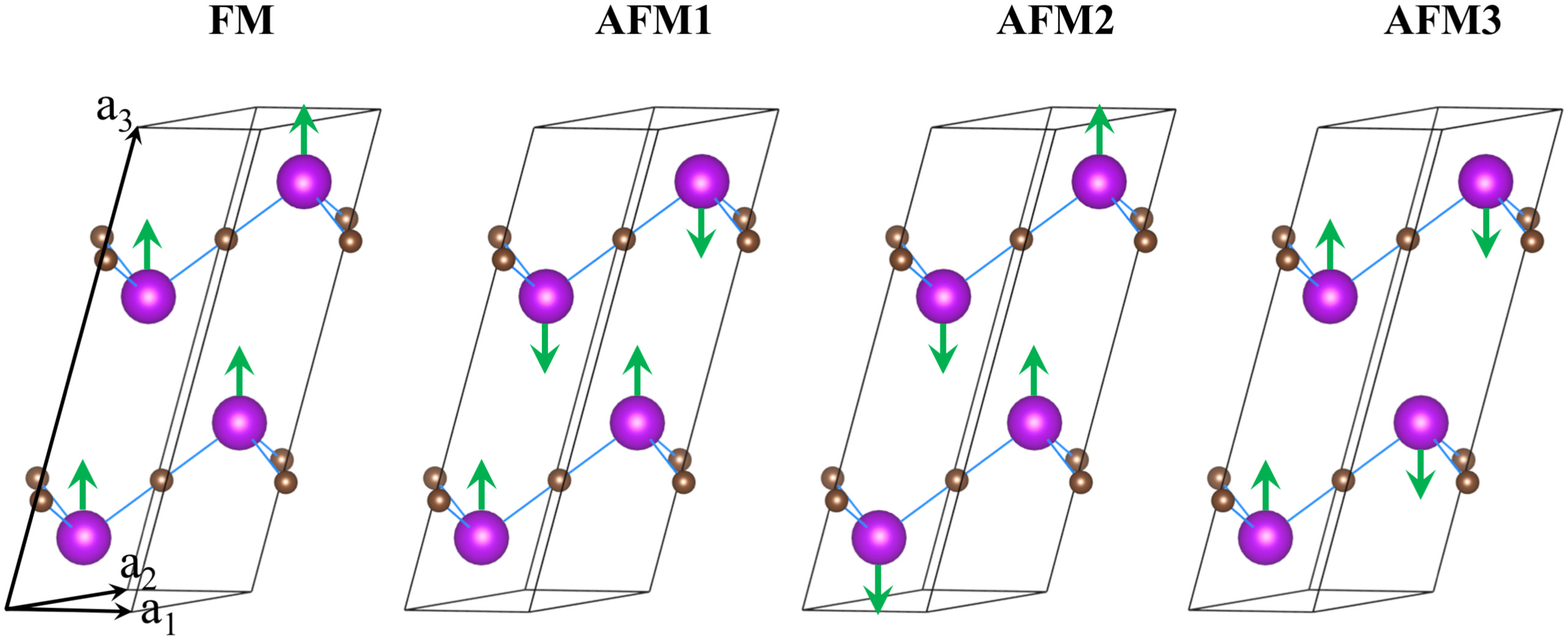}
\caption{Ferromagnetic (FM) and antiferromagnetic (AFM) configurations of bulk Gd$_2$C. Three different AFM configurations AFM$_1$, AFM$_2$, and AFM$_3$ are considered. Here, the arrows indicate up and down spins. The AFM$_1$, AFM$_2$, and AFM$_3$ configurations are less stable than the FM one by 48.6, 84.9, and 84.9 meV per Gd atom, respectively. We estimate the Curie temperature ($T_{\rm c}$) by using the mean-field approximation with  $T_{\rm c}$ =2$\Delta$$E_{\rm AFM-FM}$/3$k_{\rm B}$, where $k_{\rm B}$ is the Boltzmann constant and $\Delta$$E_{\rm AFM-FM}$ is the energy difference between AFM$_1$ and FM. The estimated value of  $T_{\rm c}$ is ${\sim}$377 K, in good agreement with the experimental value of 350 K.}
\end{figure}

\vspace{1.2cm}

{\bf 7. Electron doping in Gd$_2$C }
\begin{figure}[ht]
\includegraphics[width=10cm]{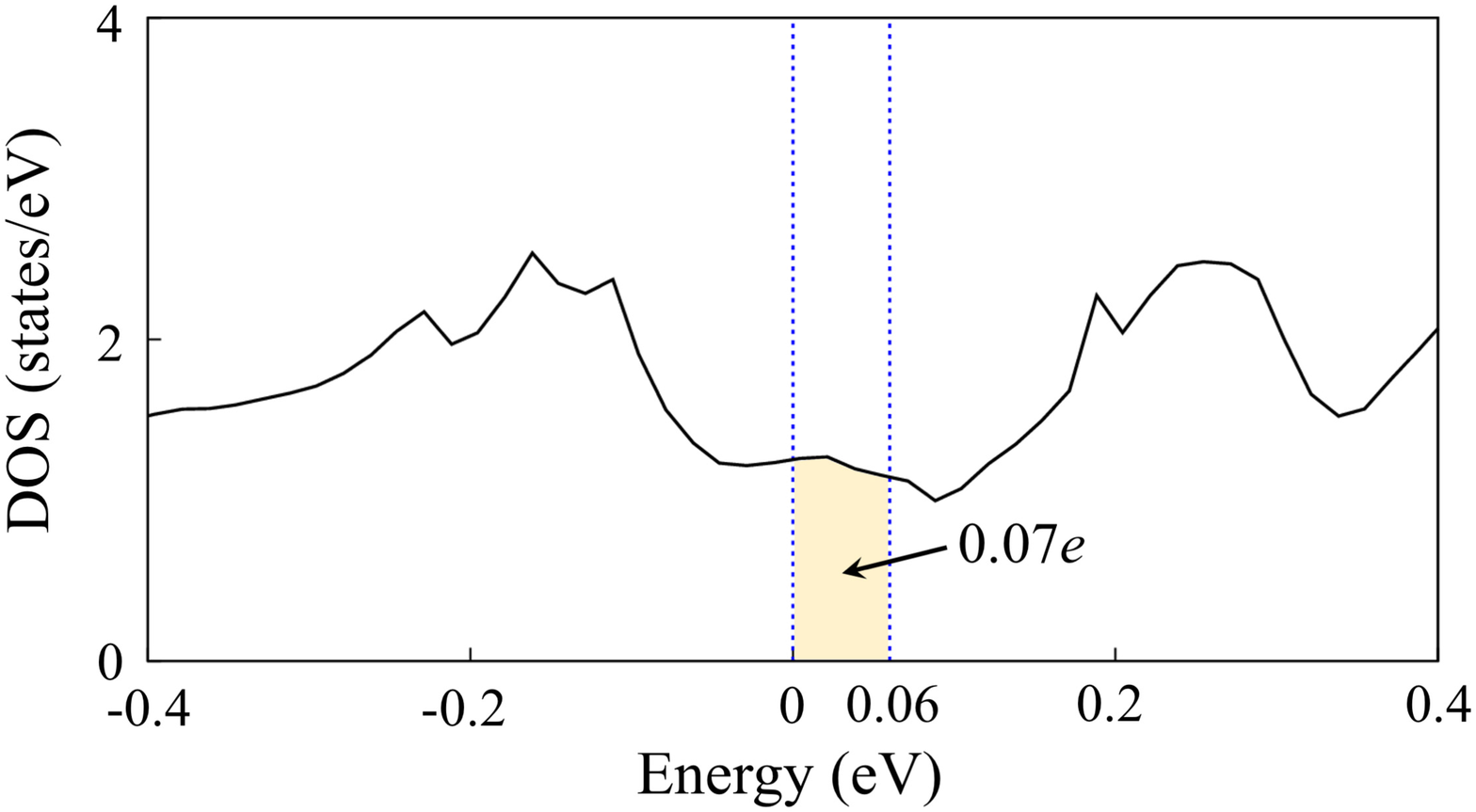}
\caption{DOS of Gd$_2$C, obtained using the DFT+U calculation with the inclusion of SOC. The number of electrons between $E_{\rm F}$ and $E_{\rm F}$+0.06 eV is ${\sim}$0.07$e$ per primitive unit cell.}
\end{figure}

\newpage

{\bf 8. Band structures and WNLs of Tb$_2$C and Dy$_2$C}
\begin{figure}[ht]
\includegraphics[width=15cm]{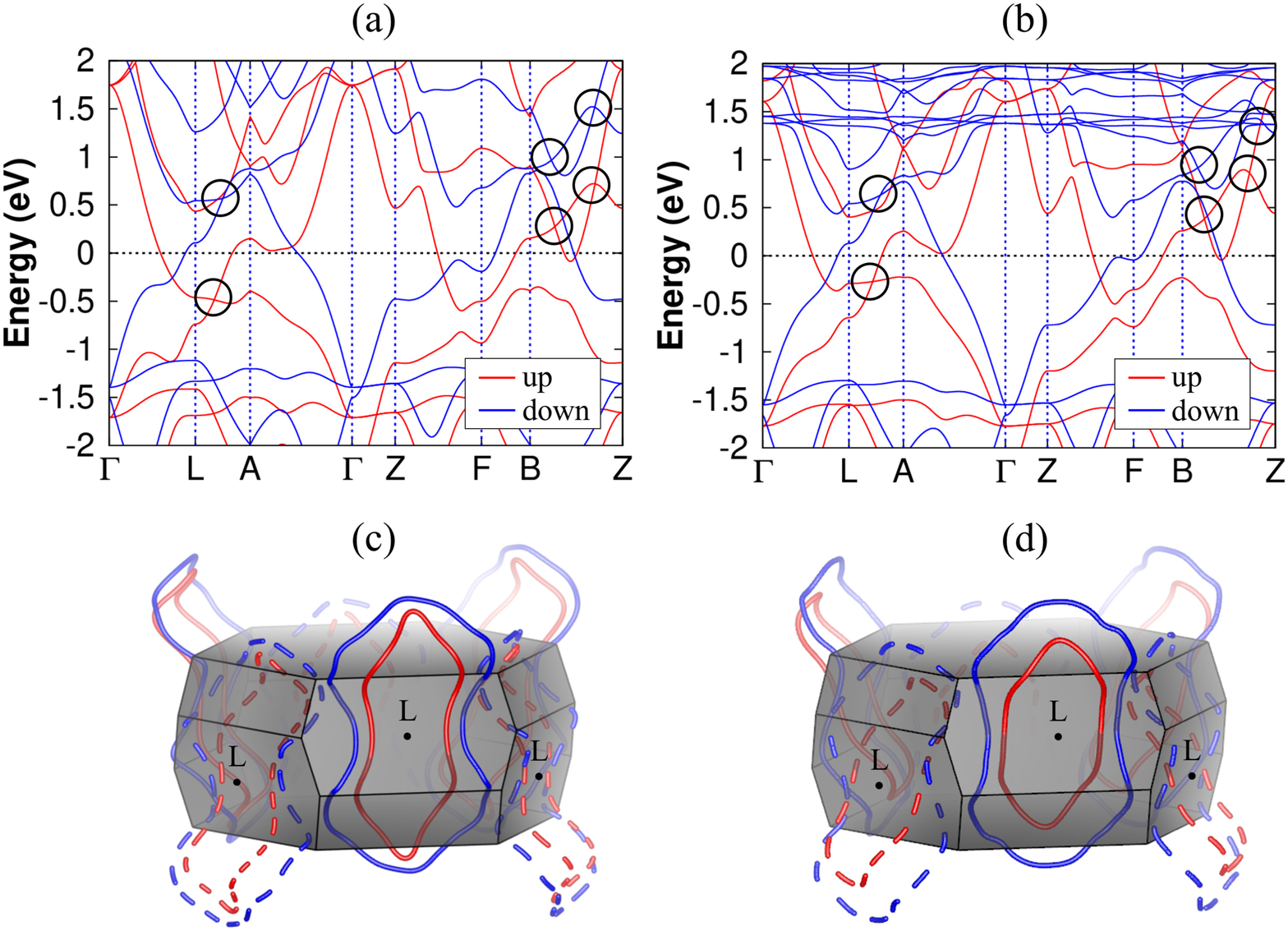}
\caption{Calculated band structures of bulk (a) Tb$_2$C and (b) Dy$_2$C using the DFT+U calculations with-out including SOC. The spinful WNLs of Tb$_2$C and Dy$_2$C, associated with the crossing points along the L-A line, are displayed in (c) and (d), respectively. The dashed lines in (c) and (d) represent the identical WNLs with modulo the corresponding reciprocal vectors.}
\end{figure}

\end{flushleft}

\vspace{3.2cm}

{\bf Table SI. Magnetic anisotropic energy (MAE) of Gd$_2$C}
\begin{table}[ht]

\renewcommand\arraystretch{2}
%\centering
%\begin{ruledtabular}
%\begin{tabular}{|p{2cm}|p{4cm}|p{4cm}|c|c|c|}
%\begin{tabular}{p{4cm}|ccc}
%\begin{tabular}{p{4cm}<{\centering}|ccc}
\begin{tabular}{p{4cm}<\centering p{4cm}<\centering p{4cm}<\centering }
\hline
$\theta$($^{\circ}$) &45 & 90 \\ 
 \hline
$E_{\rm MAE}$ (meV) & 0.06 & 0.12 \\
\hline
\end{tabular}
\caption{Calculated MAE values (in meV per primitive unit cell) of Gd$_2$C with respect to the magnetic easy axis in the $z$ direction. Here, the angle $\theta$ = 90$^{\circ}$ is along the $x$ axis}
\vspace{0.2cm}
%\end{ruledtabular}
\end{table}

\newpage

{\bf Table SII. Positions and energies of Weyl nodes}
\begin{table}[ht]
\renewcommand\arraystretch{2}
\begin{tabular}{p{2.8cm}<\centering|p{3.5cm}<\centering|p{3.5cm}<\centering|p{3.5cm}<\centering|p{2.6cm}<\centering}
\hline
Weyl node & $k_x$ & $k_y$ &  $k_z$ & Energy (eV)\\
\hline
W$_{1}^{+}$(W$_{1}^{-}$) & -0.877 (0.877)& 0 (0)&-0.383 (0.383) &-0.898 \\
\hline
W$_{2}^{+}$(W$_{2}^{-}$) &-0.869 (0.869) & 0 (0)& -0.315 (0.315)&0.391 \\
\hline
W$_{3}^{+}$(W$_{3}^{-}$) &0.802 (-0.802) & 0.101 (-0.101)&0.475 (-0.475) &-0.839 \\
\hline
W$_{3'}^{+}$(W$_{3'}^{-}$) &0.802 (-0.802) & -0.101 (0.101)&0.475 (-0.475) &-0.839 \\
\hline
W$_{4}^{+}$(W$_{4}^{-}$) &-0.962 (0.962) &-0.314 (0.314) & 0.106 (-0.106)&-0.458\\
\hline
W$_{4'}^{+}$(W$_{4'}^{-}$) &-0.962 (0.962)&0.314 (-0.314) &0.106 (-0.106)&-0.458 \\
\hline
W$_{5}^{+}$(W$_{5}^{-}$) &-0.898 (0.898) &-0.436 (0.436) &-0.452 (0.452) &0.993 \\
\hline
W$_{5'}^{+}$(W$_{5'}^{-}$) &-0.898 (0.898) & 0.436 (-0.436)& -0.452 (0.452)&0.993 \\
\hline
W$_{6}^{+}$(W$_{6}^{-}$) & -0.139 (0.139)&0.241 (-0.241) &-0.355 (0.355) &0.765 \\
\hline
W$_{6'}^{+}$(W$_{6'}^{-}$) &-0.139 (0.139) & -0.241 (0.241)&-0.355 (0.355) &0.765 \\
\hline
W$_{7}^{+}$(W$_{7}^{-}$) &0.279 (-0.279) & 0 (0)& -0.355 (0.355)& 0.765\\
\hline
W$_{8}^{+}$(W$_{8}^{-}$) &0.742 (-0.742) &-0.315 (0.315) &0.160 (-0.160) & 0.125\\
\hline
W$_{8'}^{+}$(W$_{8'}^{-}$) &0.742 (-0.742) &0.315 (-0.315) &0.160 (-0.160) &0.125 \\
\hline
W$_{9}^{+}$(W$_{9}^{-}$) &-0.259 (0.259) & -0.259 (0.259)&0.339 (-0.339) &1.820 \\
\hline
W$_{9'}^{+}$(W$_{9'}^{-}$) & -0.259 (0.259)&0.259 (-0.259) &0.339 (-0.339) &1.820 \\
\hline
W$_{10}^{+}$(W$_{10}^{-}$) & 0.212 (-0.212)& -0.177 (0.177)& 0.339 (-0.339)&1.820 \\
\hline
W$_{10'}^{+}$(W$_{10'}^{-}$) & 0.212 (-0.212)&0.177 (-0.177) &0.339 (-0.339) &1.820 \\
\hline
W$_{11}^{+}$(W$_{11}^{-}$) &0.046 (-0.046) & -0.274 (0.274)&0.339 (-0.339) &1.820 \\
\hline
W$_{11'}^{+}$(W$_{11'}^{-}$) &0.046 (-0.046) &0.274 (-0.274) &0.339 (-0.339) &1.820 \\
\hline
W$_{12}^{+}$(W$_{12}^{-}$) & 0.666 (-0.666)& 0.483 (-0.483)& -0.263 (0.263)&1.211 \\
\hline
W$_{12'}^{+}$(W$_{12'}^{-}$) & 0.666 (-0.666)& -0.483 (0.483)&-0.263 (0.263) &1.211 \\
\hline
W$_{13}^{+}$(W$_{13}^{-}$) & 0.255 (-0.255)& -0.441 (0.441)& 0.503 (-0.503)&-0.255 \\
\hline
W$_{13'}^{+}$(W$_{13'}^{-}$) &0.255 (-0.255) & 0.441 (-0.441)& 0.503 (-0.503)&-0.255 \\
\hline
W$_{14}^{+}$(W$_{14}^{-}$) & -0.508 (0.508)& 0 (0)& -0.503 (0.503)&-0.255\\
\hline

\end{tabular}
\caption{The positions of all twenty-four pairs of Weyl nodes in momentum space, together with their energies. The positions ($k_x$, $k_y$, $k_z$) are in units of {\AA}$^{-1}$. Energies are relative to $E_{\rm F}$.}
\end{table}

\end{document}